\documentclass[11pt]{article}

\usepackage{siunitx}
\usepackage{amssymb, amsmath, color,graphicx, url}
\usepackage{natbib}
\usepackage{refcount}

\title{Bayesian method for inferring the impact of geographical distance on intensity of communication}
\author{}

\begin{document}	
	\maketitle 
	
	Fei Li\footnote{\label{note1}Department of Biostatistics, Harvard School of Public Health, Boston MA 02115 USA} \footnote{\label{note3}Funding info: NIAID R37 51164, R01 AI112339}, Jukka-Pekka Onnela\footnotemark[\getrefnumber{note1}] \footnotemark[\getrefnumber{note3}], Victor DeGruttola\footnotemark[\getrefnumber{note1}] \footnotemark[\getrefnumber{note3}]  \footnote{Corresponding author: Victor DeGruttola, degrut@hsph.harvard.edu} 
	
	\begin{abstract}
	Both theoretical models and empirical findings suggest that the intensity of communication among groups of people declines with their degree of geographical separation. There is some evidence that rather than decaying uniformly with distance, the intensity of communication might decline at different rates for shorter and longer distances. Using Bayesian LASSO for model selection, we introduce a statistical model for estimating the rate of communication decline with geographic distance that allows for discontinuities in this rate. We apply our method to an anonymized mobile phone communication dataset. Our results are potentially useful in settings where understanding social and spatial mixing of people is important, such as in cluster randomized trials design.
	
	\paragraph{keywords} Bayesian LASSO; Communication intensity; Geographic constraint
	\end{abstract}

	\section{Problems Statement and Motivation}
	Observations of one way of communication are generally informative about the use of others \citep{eagle2008mobile, wang2011human, hawelka2014geo}. For example, people who speak one the phone frequently are also likely to interact in person \citep{eagle2009inferring}. For researchers studying infectious diseases, such as HIV or Malaria, the structure of social interactions in a population can provide valuable insights into how viruses are transmitted among members of that population \citep{gregson2002sexual, jones2003assessment,helleringer2007sexual}. Because traditional surveys are resource intensive and scale poorly, mobile phone data, or more specifically call detail records (CDRs), have emerged as an alternative for inferring the structure of underlying interpersonal interactions \citep{onnela2007structure, buckee2013mobile, tatem2014integrating}.
	
	Although user interactions on the mobile phone network are not limited by geography, users themselves are subject to spatial constraints that restrict the locations they may frequent and therefore influence their overall interpersonal and mobile phone communication patterns. Individual-level analysis in \cite{sailer2012social} demonstrated a relationship between spatial configuration of offices and social connections among employees. Overlap of geographical space and information flow network is discussed in \citet{ter2009applying} from a perspective of the spread of innovation and knowledge. The effect of geographic restrictions may differ for locations in different regions. For example, \cite{lambiotte2008geographical} and \cite{expert2011uncovering} found that in Belgium, cell phone users communicate mostly within language-specific network communities \citep{porter2009communities} of French and Flemish speakers. In another example, \cite{wang2013human} showed that contact patterns between individuals with respect to disease transmission are location-specific. Potential overlap of the geographical and social networks on the topological level has also been explored. The connection between local network topology and tie strength is found to be consistent with the so-called weak-ties hypothesis \citep{granovetter1973strength} in \cite{onnela2007structure}. However, geographical and community centrality were not found to be related in \cite{onnela2011geographic}.

	In this study, we investigate the impact of spatial distance on cell phone communication using a statistical approach. Our choice of model is guided by the observation that the intensity of communication among groups of people tends to decay with geographical distance and, further, the rate of decay in intensity appears to differ between short and long distances. To incorporate this feature, we allow for the existence of a break-point in the relationship between communication intensity and spatial distance. As the structure of electronic communication, human mobility and travel, and in-person social interactions are all related, we make use of existing methods and models in these areas. Some of the most widely studied models in these fields are the gravity model \citep{krings2009urban, lambiotte2008geographical,balcan2009multiscale,noulas2012tale,csaji2013exploring}, the radiation model \citep{simini2012universal}, and the rank-based friendship model \citep{liben2005geographic}. Both the radiation model and the rank-based friendship model make explicit mechanistic assumptions regarding the effect of distance and population sizes, and these models focus on prediction. The gravity model is simpler and ignores the geographical distribution of the population, as it uses only the source and destination population sizes and the spatial distance between them.
	
	We extend the gravity model by relaxing the assumption of homogeneity in distance effects; its unsatisfactory performance in prediction compared with the radiation model shown in \cite{simini2012universal} is mainly due to the assumption of an identical decay rate for all distances. We explicitly incorporate the potential for heterogeneity of distance effects into our model, and we also provide an estimate and an interval for the break-point between short and long distances.  
	
	This paper is organized as follows. Section 2 and 3 introduce notation and the models, sampling schemes, diagnosis of convergence and computational complexity. Section 4 describes the design of the simulation study and the dataset we analyze. Section 5 provides the results of the simulation study, and Section 6 presents the data analyses. We conclude with a discussion in Section 7.

	\section{Notation and Theory}
	Our starting point to investigating the relationship between spatial distance and communication intensity is the so-called gravity model. Using the notation from \cite{krings2009urban}, the gravity model can be written as
	\begin{equation}
		G_{ij}=K \dfrac{m_i n_j}{r_{ij}^2},
	\end{equation}
	where $G_{ij}$ specifies the communication intensity from source location $i$ to destination location $j$, $K$ is a constant, $m_i$ is the population of the source location $i$, $n_j$ is the population of the destination location $j$, $r_{ij}$ is the distance between source $i$ and destination $j$. 
	
	In \cite{simini2012universal}, the model is extended to the following:
	\begin{equation}
		G_{ij}= \dfrac{m_i^{\alpha} n_j^{\beta}}{f(r_{ij})},
	\end{equation}
	where $f(\cdot)$ is a function that specifies the decay of $G_{ij}$ with distance $r_{ij}$, and it is usually specified as $r_{ij}^{\gamma}$. Here, we adopt the following form of the model:
	\begin{equation}
		G_{ij}= K \dfrac{m_i^{\alpha} n_j^{\beta}}{r_{ij}^{\gamma}}.
	\end{equation}
	
	Taking the logarithm of this yields
	\begin{equation}
		\label{grav1}
		\log(G_{ij})= \log(K) + \alpha \log(m_i) + \beta \log(n_j) - \gamma \log(r_{ij}).
	\end{equation}

	\section{A Bayesian approach}
	\subsection{Model}
	We extend the gravity model shown in Equation \ref{grav1} in the following way:
	\begin{equation}
		\begin{split}
			Y_{ij} & = \mu+ \beta_1 \log(n_i) + \beta_2 \log(n_j) + \beta_{3,i} \log(d_{ij}) + \beta_{4,i} (\log(d_{ij})-\theta_{i})_+ + \epsilon_{ij},\\
			& i,j=1,...,S, j \neq i,
		\end{split}
	\end{equation}
	where $Y_{ij}=g(G_{ij})$ and $g(\cdot)$ is a transformation function, in the gravity model, $g(\cdot)=\log(\cdot)$; $\mu$ is the intercept; $\theta_{i}$ represents the location of the break point measured on the logarithmic scale for communication initiated from location $i$; $\beta_{3,i}$ represents the distance effect before break point $\theta_{i}$; and $\beta_{4,i}$ specifies the difference of distance effect before and after the break point. When $\beta_{4,i}=0$, the difference is 0, i.e. the rate of decay does not change over the observed range. We denote the size of the population at location $i$ as $n_i$ and refer the model with $\beta_{4,i}$ as the \emph{full model} and the model that sets $\beta_{4,i}$ to 0 as the \emph{reduced model}. By definition, $(d_{ij}-\theta_{i})_+=(d_{ij}-\theta_{i}) I (d_{ij}>\theta_{i})$, which takes value 0 before the break point $\theta_{i}$ and $d_{ij}-\theta_{i}$ after the break point. We assume that $\epsilon_{ij} \stackrel{iid}{\sim} N(0, \sigma^2)$; $S$ denotes the number of locations, and $i$ and $j$ are indexes. This formulation provides a straightforward way to compare the two nested models with regard to the effect of distance effect; the reduced model has the constraint $\beta_{4,i}=0$. In this formulation, model selection  becomes a variable selection problem that can be achieved using a variety of methods, such as LASSO. We are also interested in estimating $\theta_i$ and quantifying its uncertainty. To achieve these goals, we employ a Bayesian approach with a Metropolis sampling block for $\boldsymbol{\theta} \equiv (\theta_1, \theta_2, ..., \theta_S)^T$ and a Bayesian LASSO block dealing with $\boldsymbol{\beta_4} \equiv (\beta_{4,1}, \beta_{4,2}, ..., \beta_{4,S})^T$.
	
	We note that the above model assumes that the full and nested models share the same intercept and population size effects --- an assumption that might not hold in practice. To address this concern, we consider two distinct settings, or cases. In what follows, \emph{case I}  refers to the setting where the assumption holds, and \emph{case II}, to the setting it does not. For the latter, we extend the model by making use of the Reversible Jump MCMC (RJMCMC) option in the \verb|blasso| function in \verb|R| package \verb!monomvn!. This approach allows for statistical inference using Bayesian LASSO. Briefly, RJMCMC sampling procedure permits a change in the model matrix based on the variable selection results from the previous iteration; the intercept and population size effects are modeled separately for the two models. We provide details in the next section.
	
	\subsection{Sampling algorithm}
	
	\subsubsection{Initial values}
	To speed up convergence and prevent the algorithm from converging to a local mode, we calculate a set of crude initial values for all the parameters as follows:
	
	1. Search through a grid over the distance range of location $i$ for $\theta_i$ and choose the grid point that maximizes the likelihood function of the crude full model $\boldsymbol{\theta^{(0)}}$. 
	
	2. For case I, the preliminary values for the parameters are obtained by linear regression treating the break points as known. Substituting in the value of $\boldsymbol{\theta^{(0)}}$ from Step 1 leads to crude parameter estimates $\mu^{(0)}$, $\boldsymbol{\beta^{(0)}} \equiv {(\beta_1^{(0)}, \beta_2^{(0)}, {\boldsymbol{\beta_{3}^{(0)}}}^T, {\boldsymbol{\beta_{4}^{(0)}}}^T)}^T$ and $\sigma^{2}_{(0)}$. For case II, we fit two models for each source location: Model 1 has a break point at $\boldsymbol{\theta^{(0)}}$ estimated in Step 1 and Model 2 has no break point. We then assign $\eta_{i}^{(0)}=1$ if Model 1 has a lower BIC than Model 2, and assign $\eta_{i}^{(0)}=0$ otherwise. We use BIC to account for the fact that Model 1 has more parameters than Model 2. Based on $\boldsymbol{\eta^{(0)}} \equiv (\eta_{1}^{(0)},\eta_{2}^{(0)},...,\eta_{S}^{(0)})^T$, we create a new corresponding model matrix, removing the column of $\beta_{4,i}$ if $\eta_{i}^{(0)}=0$, and obtain the crude parameter estimates $\boldsymbol{\mu^{(0)}, \beta^{(0)}}$ and $\sigma^{2}_{(0)}$ from linear regression. For cases where $\eta_{i}^{(0)}=0$, we assign $\beta_{4,i} = 0$.
	
	\subsubsection{Metropolis block and Bayesian LASSO}
	
	\paragraph{Case I: Assuming same intercept and population size effects across all source locations}
	With Bayesian LASSO, the model is specified as
	\begin{equation}
		\begin{split}
			Y_{ij} & = \mu+ \beta_1 \log(n_i) + \beta_2 \log(n_j) + \beta_{3,i} \log(d_{ij}) + \beta_{4,i} (\log(d_{ij})-\theta_{i})_+ + \epsilon_{ij},\\
			& \theta_i \in (\min\limits_{j} \log(d_{ij}), \max\limits_{j} \log(d_{ij})), i,j=1,...,S, j \neq i,
		\end{split}
	\end{equation}
	which can be written as $\boldsymbol{Y=}\mu \boldsymbol{1}+\boldsymbol{X\beta+\epsilon}$ using matrix notation. $\mu$ is not included in the Bayesian LASSO penalty term \citep{park2008bayesian}; $\boldsymbol{1}$ is the vector of 1s; $\boldsymbol{X}$ is the model matrix consisting of logarithmic population sizes and distances, and $\boldsymbol{\beta}$ is the vector of $\beta$s.
	
	In general, LASSO \citep{tibshirani1996regression} solves an unconstrained optimization problem subject to a given bound on the $L_1$ norm of the parameter vector that is equivalent to
	\begin{equation}
		\label{lasso_beta}
		\min\limits_{\boldsymbol{\beta}} \boldsymbol{(\tilde{Y}-X\beta)^T(\tilde{Y}-X\beta)}+\lambda \sum\limits_{j=1}^{p} |\beta_j|,
	\end{equation}
	
	where $\boldsymbol{\tilde{Y}=Y}-\mu \boldsymbol{1}$ is the centered outcome vector; $p$ is the number of parameters after excluding the intercept. In the Bayesian setting, solution to Equation \ref{lasso_beta} provides the posterior mode estimates when $\beta_j$ has i.i.d. double exponential priors. As explained in \cite{park2008bayesian}, conditional double exponential priors are used in the formulation to avoid multiple modes. They can be expressed hierarchically as
	\begin{equation}
		\begin{split}
			& \boldsymbol{Y}|\mu, \boldsymbol{X, \beta}, \sigma^2 \sim N(\mu \boldsymbol{1}+\boldsymbol{X\beta}, \sigma^2 \boldsymbol{I}),\\
			& \boldsymbol{\beta}|\tau_1^2, ..., \tau_p^2, \sigma^2 \sim N(\boldsymbol{0}, \sigma^2 \boldsymbol{D_r}) \text{, where } \boldsymbol{D_r}=\text{diag}(\tau_1^2, ..., \tau_p^2),\\
			& \sigma^2, \tau_1^2, ..., \tau_p^2 \sim \pi(\sigma^2) d\sigma^2 \prod_{j=1}^{p} \dfrac{\lambda^2}{2} e^{-\lambda^2 \tau_j^2/2} d\tau_j^2 \text{ , } \sigma^2, \tau_1^2, ..., \tau_p^2 >0.
		\end{split}
	\end{equation}
	
	The entire sampling procedure is available using function \verb|blasso| in \verb|R| package \verb|monomvn| with the option for RJMCMC specified as False. To incorporate a Metropolis block for break point estimation, we alternate between the Metropolis and Bayesian LASSO blocks. Validity of this approach is established by regarding it as two components of a Gibbs sampling algorithm. In summary, conditional on break points, our problem is one of a variable selection; conditional on other parameters, break point sampling is a straightforward application of a Metropolis algorithm.

	Thus after obtaining the initial values $\boldsymbol{\mu^{(0)}, \beta^{(0)}, \theta^{(0)}}$ and $\sigma^{2}_{(0)}$, we proceed as follows:

	1. At iteration $t$ for each source location $i$, update break point $\theta_{i}^{(t+1)}$ using Metropolis algorithm with a normal proposal $N(\theta_i^{(t)}, \sigma^2_{\theta})$.  The range of $\theta_i$ is determined empirically from data, i.e., the posterior likelihood of $\theta_i$ has an indicator function term in the product that is 0 if the proposed $\theta_{i}^{(t+1)}$ is out of the observed empirical log-distance range, thereby assuring that any out-of-range proposal will be rejected.
	
	2. For each location $i$, if there are fewer than 5\% of data points on either side of $\theta_{i}^{(t+1)}$ for the subset of data, i.e., $\boldsymbol{Y_{i}}$, we consider it to be on the boundary, specify $\beta_{4,i}^{(t+1)}=0$, and remove it from the model in the next estimation step. We denote the number of locations belonging to the boundary sets as $b^{(t+1)}$. 
	
	3. Create the corresponding $s(s-1) \times (2+2s-b^{(t+1)})$ covariate matrix (intercept column is not included) based on $\boldsymbol{\theta^{(t+1)}}$. Together with the data, $\boldsymbol{\beta^{(t)}}$ (after $\beta_{4,i}^{(t+1)}=0$ are removed), $\sigma^{(t) 2}$ and $\lambda^{(t)}$, input the covariate matrix into the \verb|blasso| function for $h$ iterations (2 or more). The output intercept is $\mu^{(t+1)}$. From the output we also get $\boldsymbol{\beta^{(t+1)}}$ ($\beta_{4,i}^{(t+1)}=0$ are put back), $\sigma^{(t+1) 2}$ and $\lambda^{(t+1)}$.

	4. Repeat steps 1-3 until convergence (see below).

	\paragraph{Case II: Allowing different intercepts and population size effects for models with and without break-points}
	
	When there is evidence of the presence of break-points, we estimate these parameters separately in two different models. In this case, estimates of intercepts and population size effects depend on the set of source locations whose data contribute to the estimation in any given iteration. We denote the mean model as $\boldsymbol{\eta^{(t)}}$ for iteration $t$ to maintain consistency with the notation we introduced earlier.

	Estimation can be done using the Reversible Jump MCMC option in the \verb|blasso| function, which allows sampling from different models. In our case, different models imply different specification of zeros in $\boldsymbol{\beta_4^{(t)}}$ , and are characterized by $\boldsymbol{\eta^{(t)}}$, where $\eta_i^{(t)}=I(\beta_{4,i}^{(t)}>0)$. 
	
	RJMCMC is a general version of the Metropolis-Hastings algorithm introduced by \cite{green1995reversible}, which allows transitions between different states or models of different dimensions. In RJMCMC, trans-dimensional moves are possible through dimension matching by augmenting the parameter vector with a random component. The difference compared to the usual Metropolis-Hastings procedure is the addition of a Jacobian term in the acceptance probability. A thorough review of RJMCMC with more recent comments can be found in \cite{green2009reversible}.

	Use of RJMCMC yields the following sampling scheme:

	1. The first two steps are the same as in case I: At iteration $t$, for each source location $i$, update break point $\theta_{i}^{(t+1)}$ using Metropolis algorithm with a normal proposal $N(\theta_i^{(t)}, \sigma^2_{\theta})$. For each location $i$, if there are fewer than 5\% of data points on either side of $\theta_{i}^{(t+1)}$ for $\boldsymbol{Y_{i}}$, we specify $\beta_{4,i}^{(t+1)}=0$ and	 remove it from the model in the next estimation step.
	
	2. Conditional on $\boldsymbol{\theta^{(t+1)}}$, create the $s(s-1) \times (5+2s-b^{(t+1)})$ covariate matrix (intercept column is not included). Data from each source location contribute to their own group's estimation of intercept and population size effects, which depends on $\boldsymbol{\eta_i^{(t)}}$. All data and parameter values from the previous iteration $t$ (including $\sigma^{(t) 2}$ and $\lambda^{(t)}$) are used in the \verb|blasso| function with RJMCMC for 3 iterations. 3 is the minimum number of iterations to avoid the situation in which zeros in the previous iteration are carried forward. 
	
	3. From Step 2 we get the updated $\boldsymbol{\beta^{(t+1)}}, \sigma^{(t+1) 2}, \mu^{(t+1)}$ and $\lambda^{(t+1)}$. Now update the $\boldsymbol{\eta^{(t+1)}}$: $\eta_i^{(t+1)}=1$ if $\beta_{4,i}^{(t+1)}>0$; otherwise 0.  
	
	4. Repeat steps 1-3 until convergence.

	\subsection{Diagnostics of convergence}
	
	The usual diagnostic framework for Bayesian LASSO \citep{gelman1992inference, brooks1998general,gelman2014bayesian} includes trace plots for different chains and calculation of the {\it Potential Scale Reduction Factor} (PSRF). Diagnostics for RJMCMC can be developed by extending that framework to include within model and between model variations of the parameters. 
	
	We make use of the work of \cite{castelloe2002convergence} who define two PSRFs in the assessment. For a chosen parameter, $\textrm{PSRF}_1$ is the ratio between total variations $\widehat{V}$ and variation within chains $W_c$; $\textrm{PSRF}_2$ is the ratio between variations within models $W_m$ and variations within models and chains $W_{m}W_{c}$. $\widehat{V}, W_c, W_m$ and $W_{m}W_{c}$ are defined as follows:
	
	\begin{equation}
		\begin{split}
			& \widehat{V}(\theta) = \dfrac{1}{CT-1} \sum\limits_{c=1}^{C} \sum\limits_{m=1}^{M} \sum\limits_{r=1}^{R_{cm}} (\theta_{cm}^r-\overline{\theta_{..}}^.)^2, \\
			& W_c(\theta)= \dfrac{1}{C(T-1)} \sum\limits_{c=1}^{C} \sum\limits_{m=1}^{M} \sum\limits_{r=1}^{R_{cm}} (\theta_{cm}^r-\overline{\theta_{c.}}^.)^2,\\
			& W_m(\theta)= \dfrac{1}{CT-M} \sum\limits_{c=1}^{C} \sum\limits_{m=1}^{M} \sum\limits_{r=1}^{R_{cm}} (\theta_{cm}^r-\overline{\theta_{.m}}^.)^2,\\
			&W_{m}W_{c}(\theta)= \dfrac{1}{C(T-M)} \sum\limits_{c=1}^{C} \sum\limits_{m=1}^{M} \sum\limits_{r=1}^{R_{cm}} (\theta_{cm}^r-\overline{\theta_{cm}}^.)^2,\\
		\end{split}
	\end{equation}
	where $\theta_{cm}^r, \overline{\theta_{..}}^., \overline{\theta_{c.}}^., \overline{\theta_{.m}}^.$ and $\overline{\theta_{cm}}^.$ are the $r^{th}$ appearance of $\theta$ in model $m$ chain $c$, mean $\theta$ across all models and chains, mean $\theta$ within chain $c$ across all models in that chain, mean $\theta$ within model m across all chains, mean $\theta$ within chain $c$ and model $m$ respectively. $R_{cm}$ is number of $\theta$ in chain $c$ model $m$. $C$ and $M$ are the number of chains and distinct models, respectively.
	
	We follow the strategy given in \cite{castelloe2002convergence} to assess convergence and, for simplicity, illustrate this approach by considering a scalar. We choose $\sigma^2$, the variance of the error terms, for this illustration, as its interpretation remains the same across the models. Each chain is divided into batches of equal length. A sequence of $\textrm{PSRF}_1$ and $\textrm{PSRF}_2$ is calculated for each batch. A desirable result is that the two quantities move toward 1 as the iteration proceeds. In the simulation study below, we illustrate the use of diagnostic graphs for evaluating convergence; further details on this subject can be found in \cite{brooks1998convergence}.

	\subsection{Interpretation}
	
	Under the assumption that intercept and population size effects are identical across source locations, we obtain a sample of $\beta_{4,i}$ as well as its 95\% credible interval rather than an estimate of the probability that each source locations has a break point. Intervals that do not cover 0 imply the presence of a break point by providing evidence against the null hypothesis that the difference of the two slopes is zero. The interpretation of other parameters is straightforward. Approaches that allow variability in intercepts and population size effects yield a sample of models and their corresponding parameter estimates. For prediction, we make use of the collection of models; the estimated mean for predicted outcomes is a weighted average of the predicted outcomes of all models.

	\subsection{Computational complexity}
	
	Because of the computational burden of these methods, we consider an analysis of a subset of data. Simulation studies (Figure \ref{fig:scaleup} in Appendix) show that computation time for the Bayesian LASSO function \verb|blasso| increases sharply  as the number of locations increases. We note that the size of the covariate matrix increases at $O(s^3)$ where $s$ specifies the number of locations. \cite{efron2004least} showed that for the least angle regression formulation of the problem, the computational complexity is $O(m^3+m^2 n)$, where $m$ is the number of features and $n$ is the number of the outcomes. In our setting, the situation is even more challenging in that the number of  outcomes grows quadratically with $s$, which renders the computational complexity to be $O(s^4)$.
	
	\section{Application: Analysis of call detail records}
	We apply our method to call detail records (CDRs) for a 3-month period to study the impact of geographical distance on communication intensity. The dataset consists of daily number of calls between distinct pairs of users. Each user is represented by a unique identifier created by the operator that made the dataset available for research. The actual phone numbers were available or recoverable from the dataset. Three covariates were made available for each person: billing zip code, sex, and age, though we only use zip code in our analysis here. We aggregated the dataset in two ways. First, we aggregated the daily call counts over the 3-month period, resulting in a single call count for each distinct pair of users. We distinguish between the caller and the receiver, so the count for each pair is directed. Second, we aggregated the data from the level of individuals to the level of counties. The resulting dataset describes communication intensity for calls among the counties. There were records for a total of 2,511,035 users; 359,759 of them resided in the largest county and 136 in the smallest one. The number of calls from one county to another ranged from 0 to 266,199, with 21,016,548 calls in total. There were 2,646 distinct zip codes nested within 427 counties. The geographical location of each county was calculated by first identifying the latitude and longitude of each zip code and then taking the mean of the these coordinates over all zip codes that were nested within a given county. For each county we thus obtained the number of users residing in that county, and for each pair of counties we obtained the spatial distance between them and the number of calls made and received by users of those counties over the 3-month period. To reduce computational burden, we selected a subset of data that arose from 65 counties with the greatest number of users. The number of users in this subgroup of counties ranged from 7,879 to 359,759. The corresponding number of calls among pairs of counties ranged from 2 to 266,226.

	\section{Simulation} 
	
	We conducted the following simulations to access the performance of our models comparing with naive approaches as well as to check the effects of different tuning parameter $\sigma^2_{\theta}$. The values of the parameters in the data generation process are selected to be the estimates from the preliminary data analysis using $\sigma^2_{\theta}=0.03$. Actual geographical distances between counties are used. We assess the performance of the gravity model, the naive fit based on BIC and grid search, and the Bayesian LASSO model on scenarios with low (0.30), medium (0.38) and high (0.45) error variances ($\sigma^2$). This division is selected such that the medium scenario matches the estimates from the preliminary analyses. For each scenario, we simulate 2 data sets and apply our algorithm with 4 chains. We also evaluate the effect of the tuning parameter for the Metropolis algorithm by specifying a series of different values for it: 0.015, 0.02, 0.025, 0.03, 0.04, 0.05, 0.06, 0.08, 0.1, 0.12, 0.15, 0.2, 0.25, 0.3, 0.4, 0.6. The diagnostic graphs in Appendix show that convergence is generally achieved. We assess the model fit and the effect of the tuning parameter based on the prediction error. One hundred new datasets were generated using the same covariates and parameters for each variance category. The findings are shown in Table \ref{prederr}.
	
	\begin{table}
		\centering
		\caption{\label{prederr}Prediction error of 3 models (2 trials each).}
		\begin{tabular}{cccccccccc}
			\hline
			&& \multicolumn{8}{c}{Variance of error term $\sigma^2$} \\
			&& \multicolumn{2}{c}{0.30} &&  \multicolumn{2}{c}{0.38} &&  \multicolumn{2}{c}{0.45}  \\
			\hline
			\multicolumn{10}{l}{Gravity model}\\
			&& 0.807 & 0.807 && 0.887 & 0.887 && 0.956 & 0.956 \\
			&&&&&&&&&\\
			\multicolumn{10}{l}{Crude model based on BIC}\\
			&& 0.331 & 0.327 && 0.412 & 0.435 && 0.485 & 0.486 \\
			&&&&&&&&&\\
			\multicolumn{10}{l}{Bayesian LASSO with breakpoints}\\
			& $\sigma^2_{\theta}$ &&&&&&&&\\
			& 0.015 & 0.321 & 0.329 && 0.403 & 0.411 && 0.479 & 0.486 \\
			& 0.020 & 0.322 & 0.329 && 0.405 & 0.413 && 0.479 & 0.487 \\
			& 0.025 & 0.319 & 0.329 && 0.404 & 0.411 && 0.479 & 0.486 \\
			& 0.030 & 0.321 & 0.326 && 0.407 & 0.411 && 0.479 & 0.485 \\
			& 0.040 & 0.318 & 0.323 && 0.409 & 0.409 && 0.481 & 0.486 \\
			& 0.050 & 0.317 & 0.323 && 0.411 & 0.411 && 0.480 & 0.487 \\
			& 0.060 & 0.318 & 0.321 && 0.411 & 0.411 && 0.481 & 0.487 \\
			& 0.080 & 0.318 & 0.321 && 0.411 & 0.410 && 0.482 & 0.487 \\
			& 0.100 & 0.318 & 0.320 && 0.411 & 0.409 && 0.482 & 0.488 \\
			& 0.120 & 0.320 & 0.320 && 0.410 & 0.412 && 0.481 & 0.485 \\
			& 0.150 & 0.319 & 0.320 && 0.411 & 0.414 && 0.486 & 0.487 \\
			& 0.200 & 0.321 & 0.319 && 0.413 & 0.414 && 0.486 & 0.490 \\
			& 0.250 & 0.321 & 0.320 && 0.415 & 0.416 && 0.489 & 0.488 \\
			& 0.300 & 0.320 & 0.321 && 0.413 & 0.417 && 0.490 & 0.489 \\
			& 0.400 & 0.321 & 0.321 && 0.417 & 0.420 && 0.496 & 0.489 \\
			& 0.600 & 0.325 & 0.322 && 0.414 & 0.419 && 0.494 & 0.489 \\				  
			\hline 
		\end{tabular}
		
	\end{table}
	
	As expected, estimates based both on BIC and Bayesian LASSO model perform better than those of the gravity model with respect to prediction error in low, medium and high error variances. The choice of tuning parameter had little effect; use of 0.2 in data analysis appears reasonable as this choice leads to a mean acceptance rate for the Metropolis algorithm on break-points in the range of 20\% to 25\% \citep{gelman2014bayesian}, as shown in Table \ref{mean_acpt}. The 95\% credible interval coverages for break points also reach high values at tuning parameter 0.2. The crude model based on BIC and Bayesian LASSO estimates are comparable. An advantage of the latter is its ability to provide interval estimates on the break points and its smaller number of required parameters. These results imply that that we did not compromise predictive power because of the estimation of location of breakpoints, though Bayesian LASSO requires greater computational time. Computation time for 15,000 iterations takes around 9 to 10 hours, whereas the BIC approach requires only a few minutes.
	
	\begin{table}
		\centering
		\caption{\label{mean_acpt}Mean acceptance rate for Metropolis algorithm on break points.}
		\begin{tabular}{c|cccccccc}
			\hline 
			&  \multicolumn{8}{c}{Variance of the error terms $\sigma^2$}  \\ 
			$\sigma^2_{\theta}$ &  \multicolumn{2}{c}{0.30} &&  \multicolumn{2}{c}{0.38} &&  \multicolumn{2}{c}{0.45} \\ 
			\hline 
			0.015 & 0.544 & 0.543 && 0.550 & 0.546 && 0.561 & 0.561 \\
			0.020 & 0.518 & 0.516 && 0.522 & 0.521 && 0.527 & 0.534 \\
			0.025 & 0.493 & 0.495 && 0.501 & 0.498 && 0.509 & 0.509 \\
			0.030 & 0.471 & 0.471 && 0.482 & 0.470 && 0.495 & 0.486 \\
			0.040 & 0.442 & 0.433 && 0.443 & 0.447 && 0.463 & 0.455 \\
			0.050 & 0.410 & 0.412 && 0.420 & 0.417 && 0.439 & 0.433 \\
			0.060 & 0.388 & 0.388 && 0.394 & 0.395 && 0.410 & 0.415 \\
			0.080 & 0.341 & 0.346 && 0.359 & 0.354 && 0.373 & 0.370 \\
			0.100 & 0.300 & 0.313 && 0.322 & 0.321 && 0.338 & 0.333 \\
			0.120 & 0.277 & 0.283 && 0.292 & 0.292 && 0.308 & 0.304 \\
			0.150 & 0.243 & 0.245 && 0.254 & 0.260 && 0.273 & 0.270 \\
			0.200 & 0.194 & 0.198 && 0.208 & 0.207 && 0.222 & 0.227 \\ 
			0.250 & 0.166 & 0.167 && 0.179 & 0.173 && 0.185 & 0.189 \\
			0.300 & 0.143 & 0.144 && 0.154 & 0.156 && 0.161 & 0.162 \\
			0.400 & 0.110 & 0.110 && 0.122 & 0.118 && 0.127 & 0.129 \\
			0.600 & 0.075 & 0.073 && 0.083 & 0.084 && 0.090 & 0.088 \\
			\hline 
		\end{tabular} 
	\end{table}
	
	\begin{table}
		\centering
		\caption{\label{coverage}95\% credible interval coverage for break points.}
		\begin{tabular}{c|cccccccc}
			\hline 
			&  \multicolumn{8}{c}{Variance of the error terms $\sigma^2$}  \\ 
			$\sigma^2_{\theta}$ &  \multicolumn{2}{c}{0.30} &&  \multicolumn{2}{c}{0.38} &&  \multicolumn{2}{c}{0.45} \\ 
			\hline
			0.015 & 0.585 & 0.585 && 0.523 & 0.492 && 0.462 & 0.492 \\
			0.020 & 0.631 & 0.615 && 0.554 & 0.523 && 0.508 & 0.569 \\
			0.025 & 0.662 & 0.631 && 0.554 & 0.523 && 0.585 & 0.600 \\
			0.030 & 0.677 & 0.677 && 0.615 & 0.554 && 0.600 & 0.631 \\
			0.040 & 0.723 & 0.723 && 0.662 & 0.631 && 0.615 & 0.615 \\
			0.050 & 0.754 & 0.692 && 0.692 & 0.692 && 0.677 & 0.646 \\
			0.060 & 0.754 & 0.708 && 0.692 & 0.692 && 0.646 & 0.646 \\
			0.080 & 0.754 & 0.754 && 0.677 & 0.738 && 0.692 & 0.677 \\
			0.100 & 0.785 & 0.785 && 0.692 & 0.754 && 0.723 & 0.692 \\
			0.120 & 0.769 & 0.815 && 0.708 & 0.738 && 0.769 & 0.708 \\
			0.150 & 0.769 & 0.815 && 0.723 & 0.738 && 0.738 & 0.692 \\
			0.200 & 0.785 & 0.877 && 0.723 & 0.738 && 0.769 & 0.677 \\
			0.250 & 0.785 & 0.862 && 0.662 & 0.738 && 0.754 & 0.708 \\
			0.300 & 0.769 & 0.862 && 0.708 & 0.738 && 0.738 & 0.692 \\
			0.400 & 0.769 & 0.862 && 0.677 & 0.708 && 0.708 & 0.708 \\
			0.600 & 0.785 & 0.831 && 0.708 & 0.738 && 0.692 & 0.708 \\
			\hline 
		\end{tabular}
	\end{table}
	
	\begin{figure}
		\centering
		\includegraphics[width=0.95\linewidth]{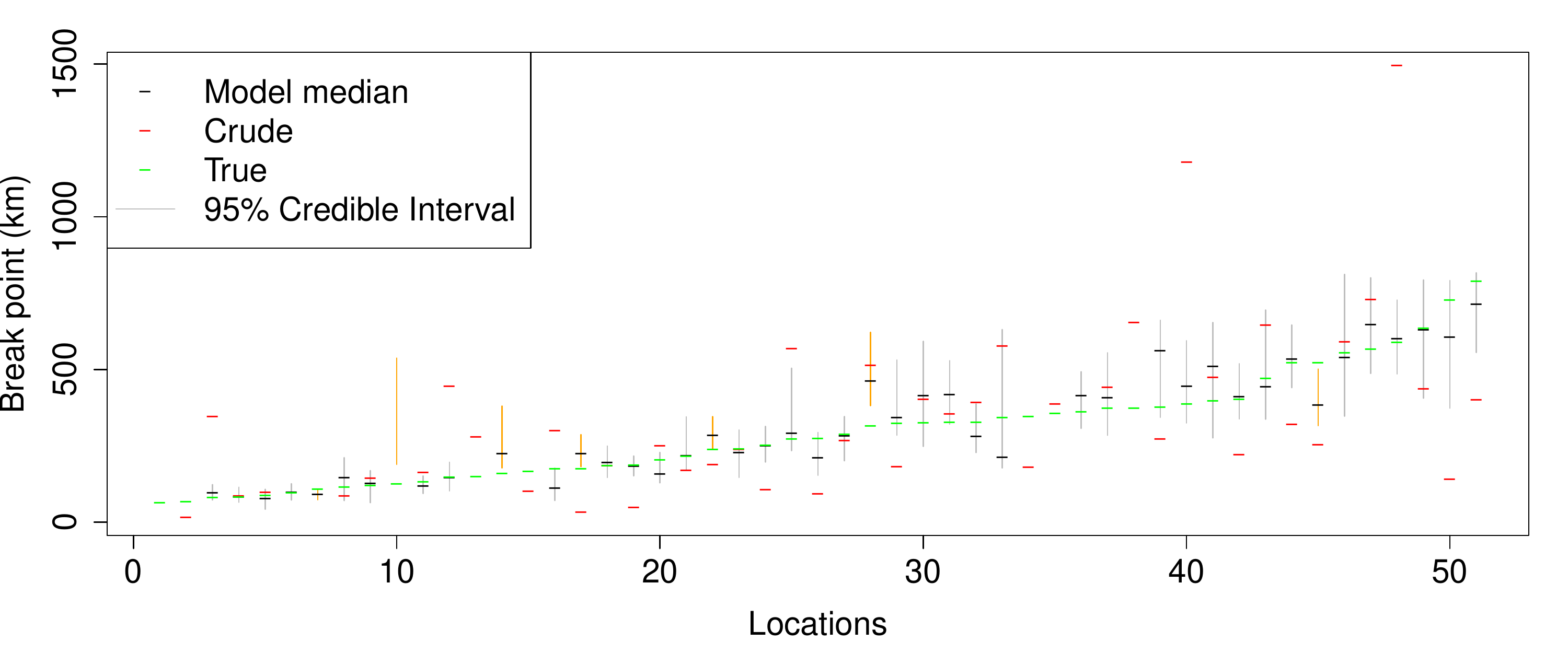}
		\includegraphics[width=0.95\linewidth]{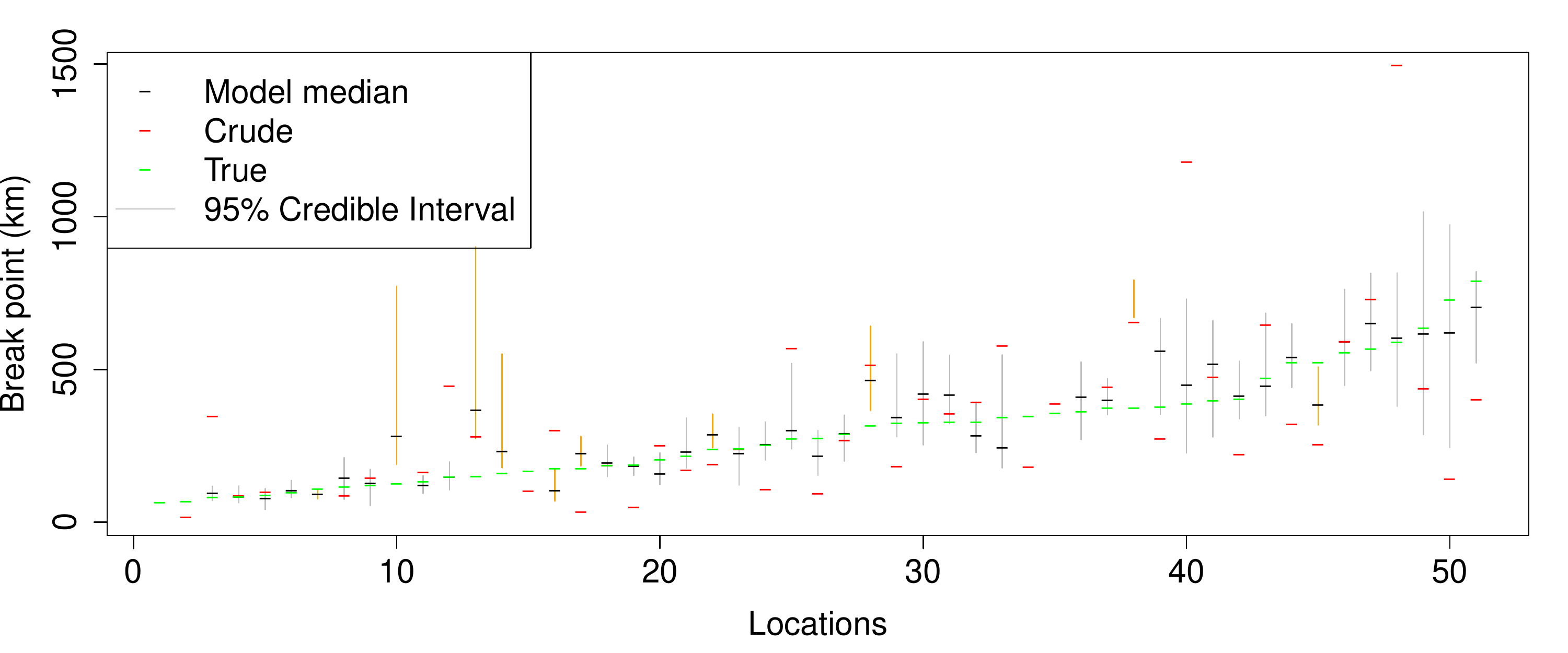}
		\includegraphics[width=0.95\linewidth]{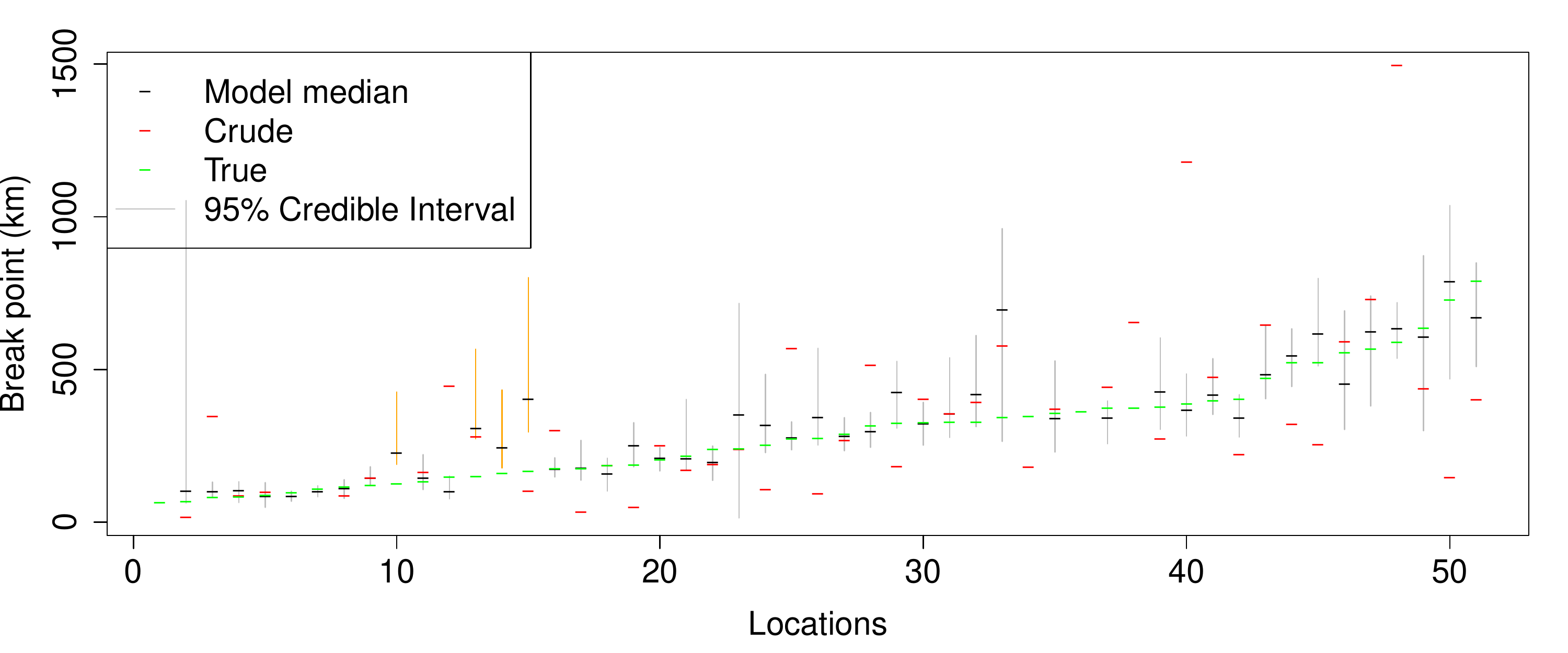}
		\caption{\label{fig:figure-1-}Estimated 95\% credible intervals of break point $\theta_{i}$ (when true break points exist) under low (top), medium (middle) and high (bottom) error variance $\sigma^2$ with tuning parameter $\sigma^2_{\theta}=0.2$; orange color of the 95\% credible interval indicates that the true value is not covered; no 95\% credible interval showing indicates none available, i.e. estimates are from model without break points.}
	\end{figure}
	
	\section{Analysis of call records data} 
	
\begin{figure}
		\centering
		\includegraphics[width=0.49\linewidth]{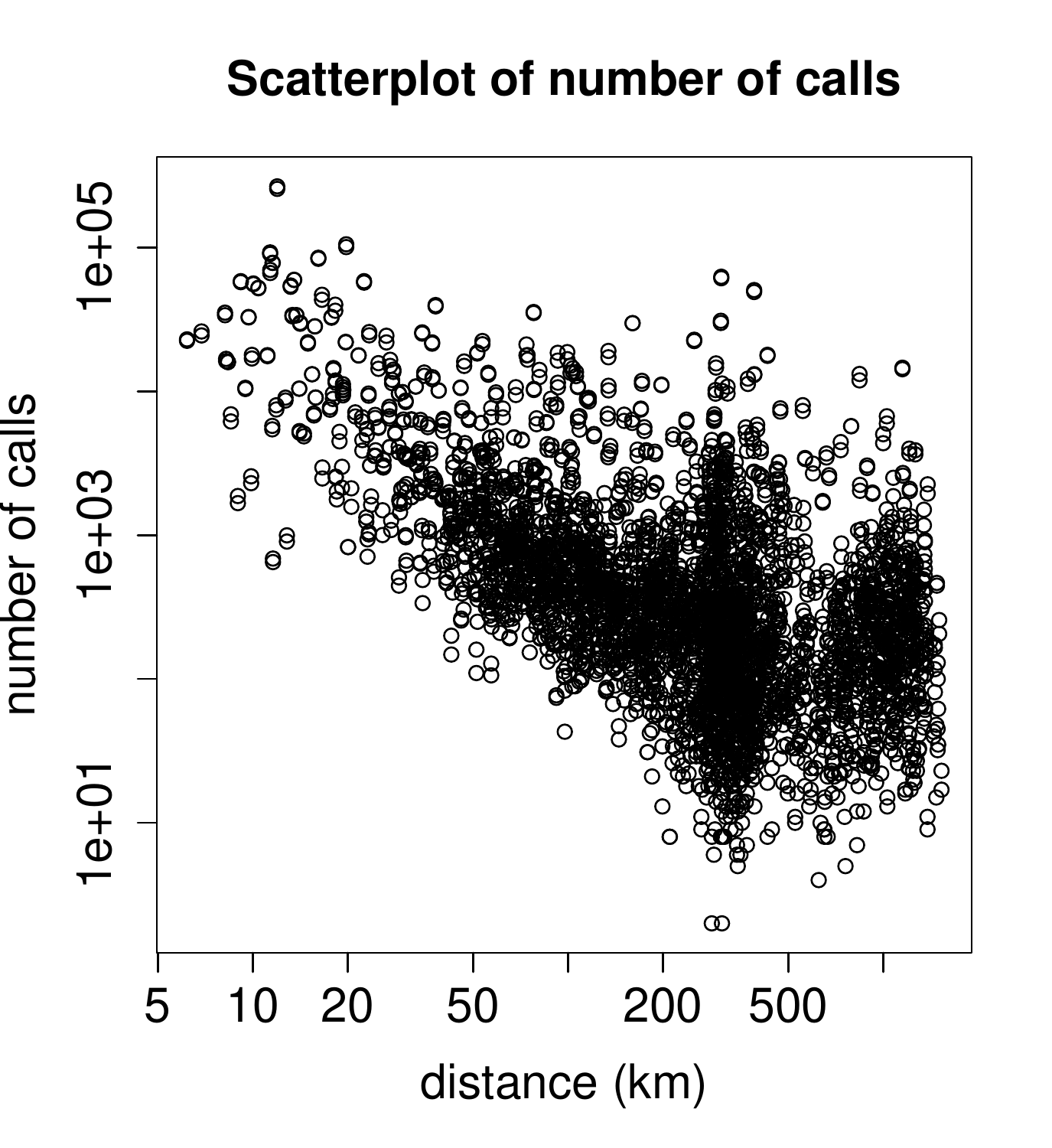}
		\includegraphics[width=0.49\linewidth]{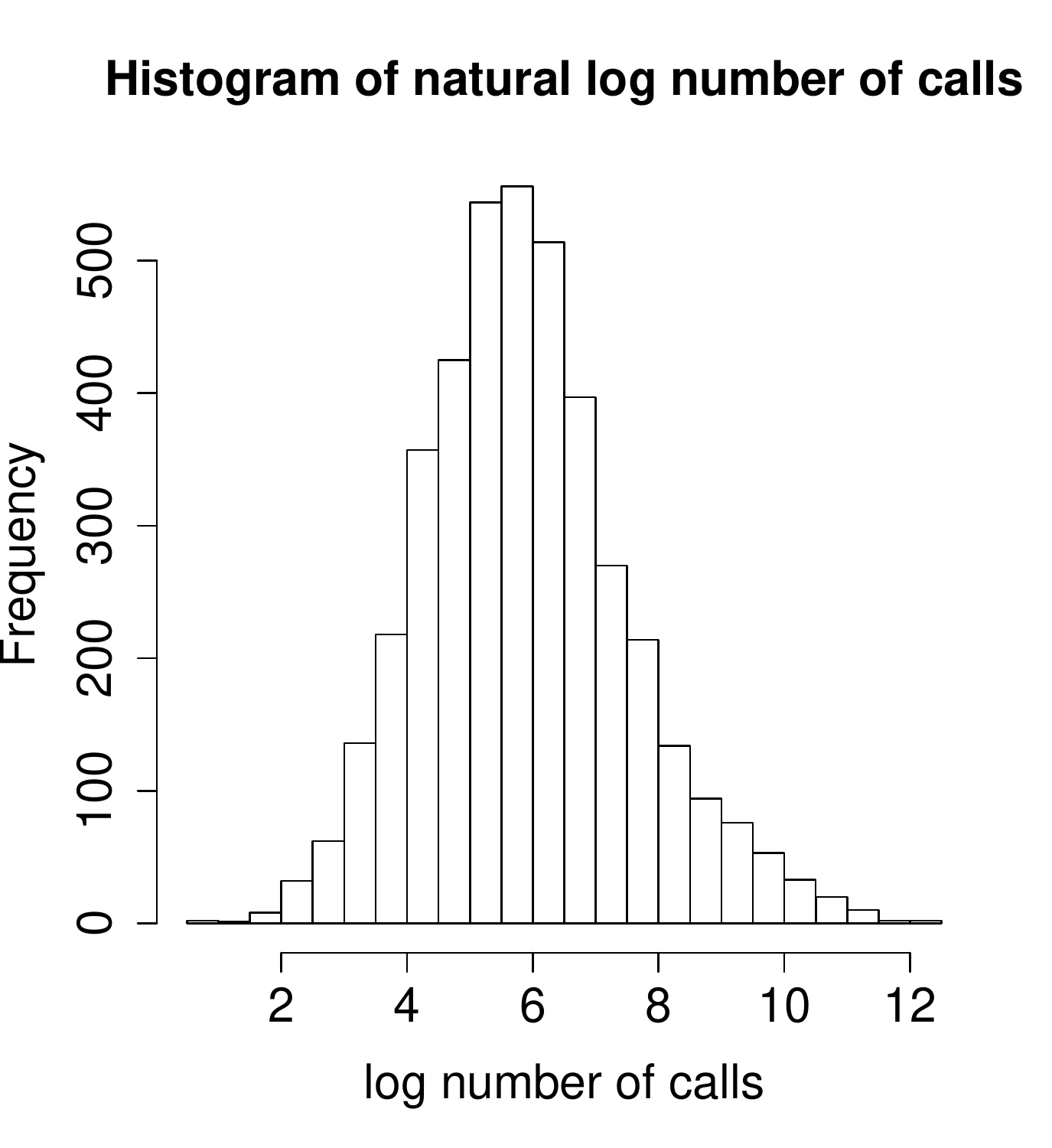}
		\caption{\label{fig:figure-2-}Left: scatter plot of natural log number of calls v.s. distances; Right: histogram of natural log number of calls.}
	\end{figure}

	
	First, we observe that Figure \ref{fig:figure-2-} is consistent with our assumptions of continuous calling intensity and normality of natural log of the number of calls.
	
	\begin{figure}
		\centering
		\includegraphics[width=0.49\linewidth]{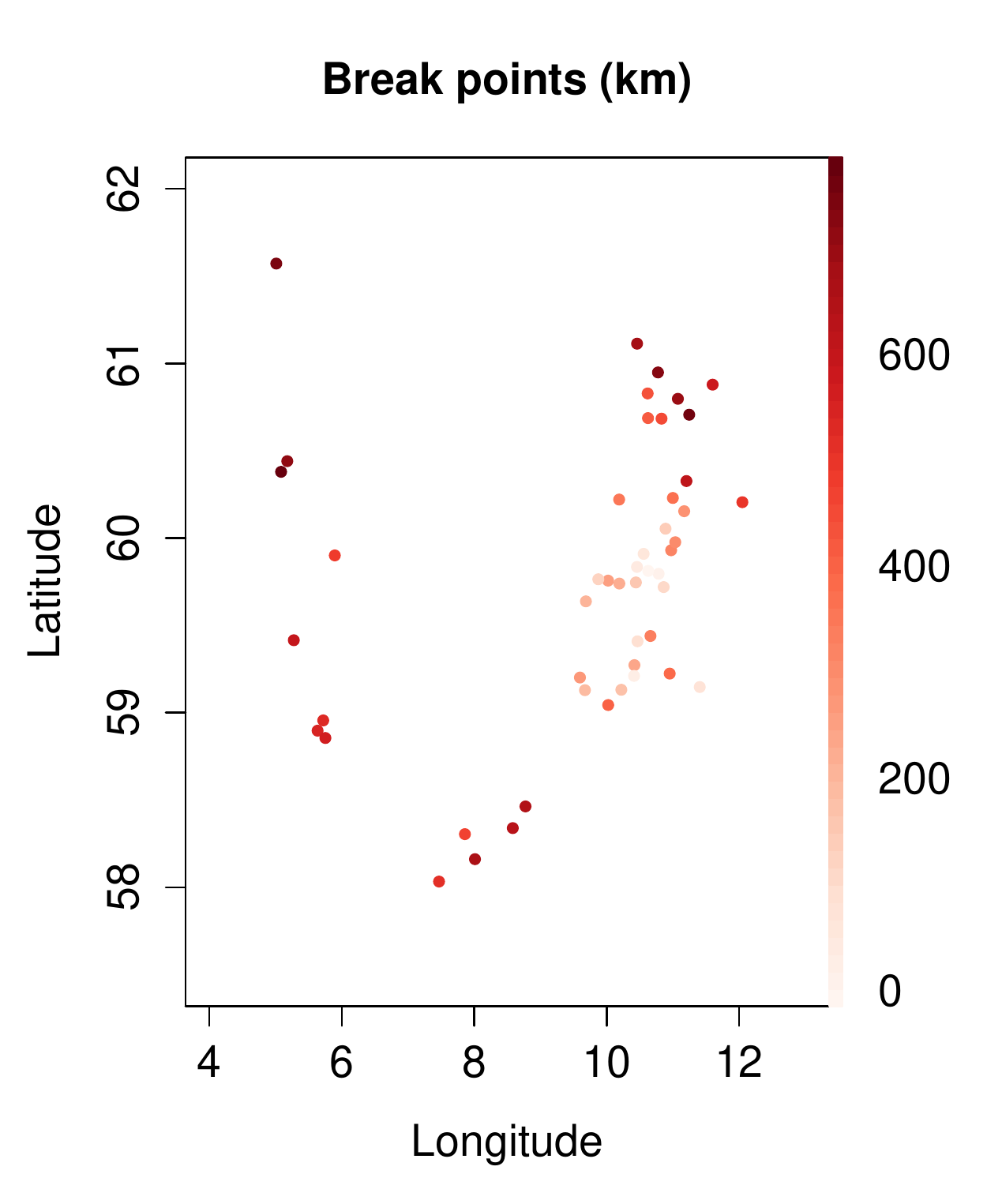}
		\includegraphics[width=0.49\linewidth]{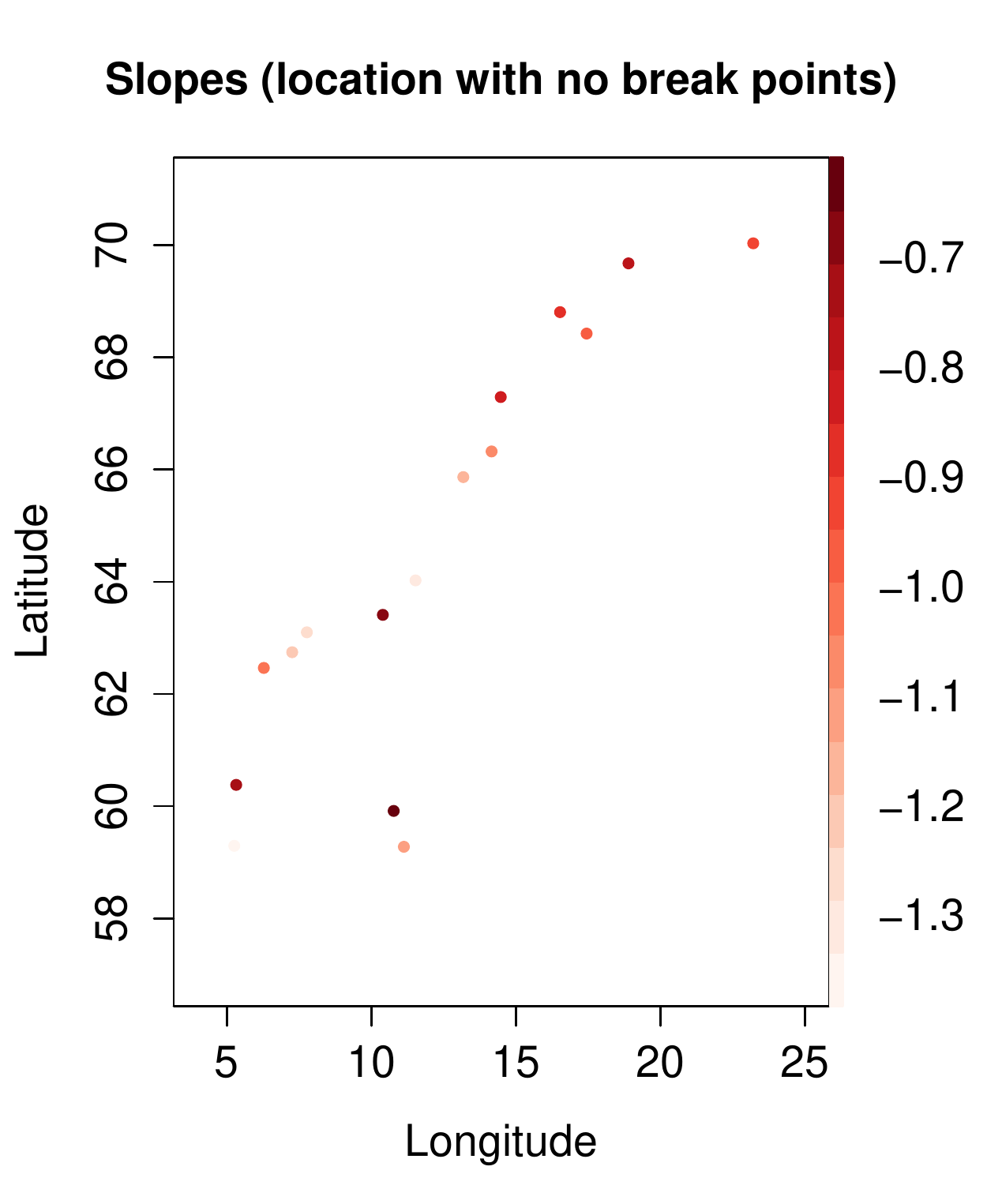}
		\includegraphics[width=1\linewidth]{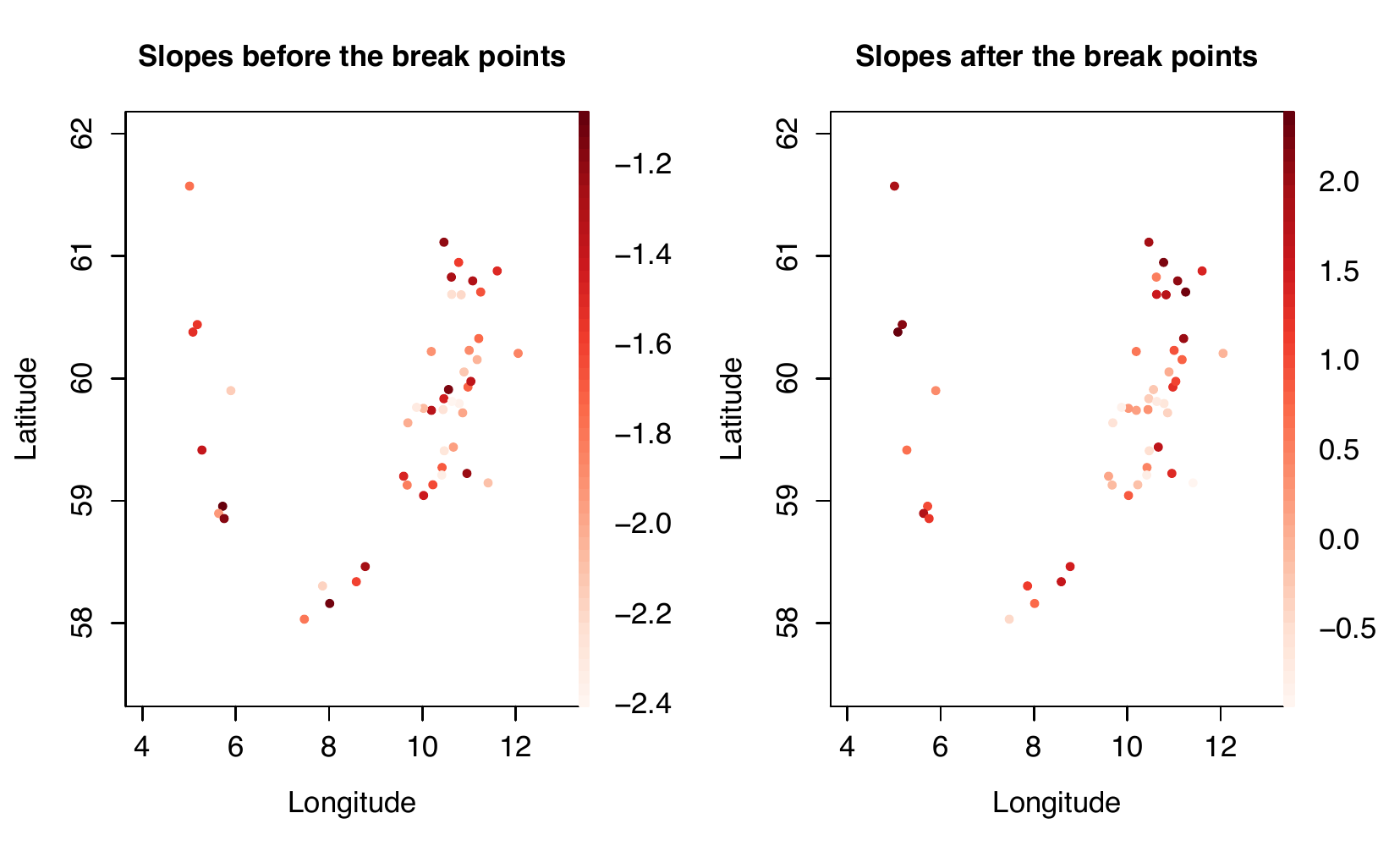}
		\caption{\label{slopes}Top left: slope estimates for locations without break points; Top right: log distance of the estimated break points for locations with break points; Bottom: Slopes estimates before and after break points for different locations.}
	\end{figure}
	\begin{figure}
		\centering
		\includegraphics[width=1\linewidth]{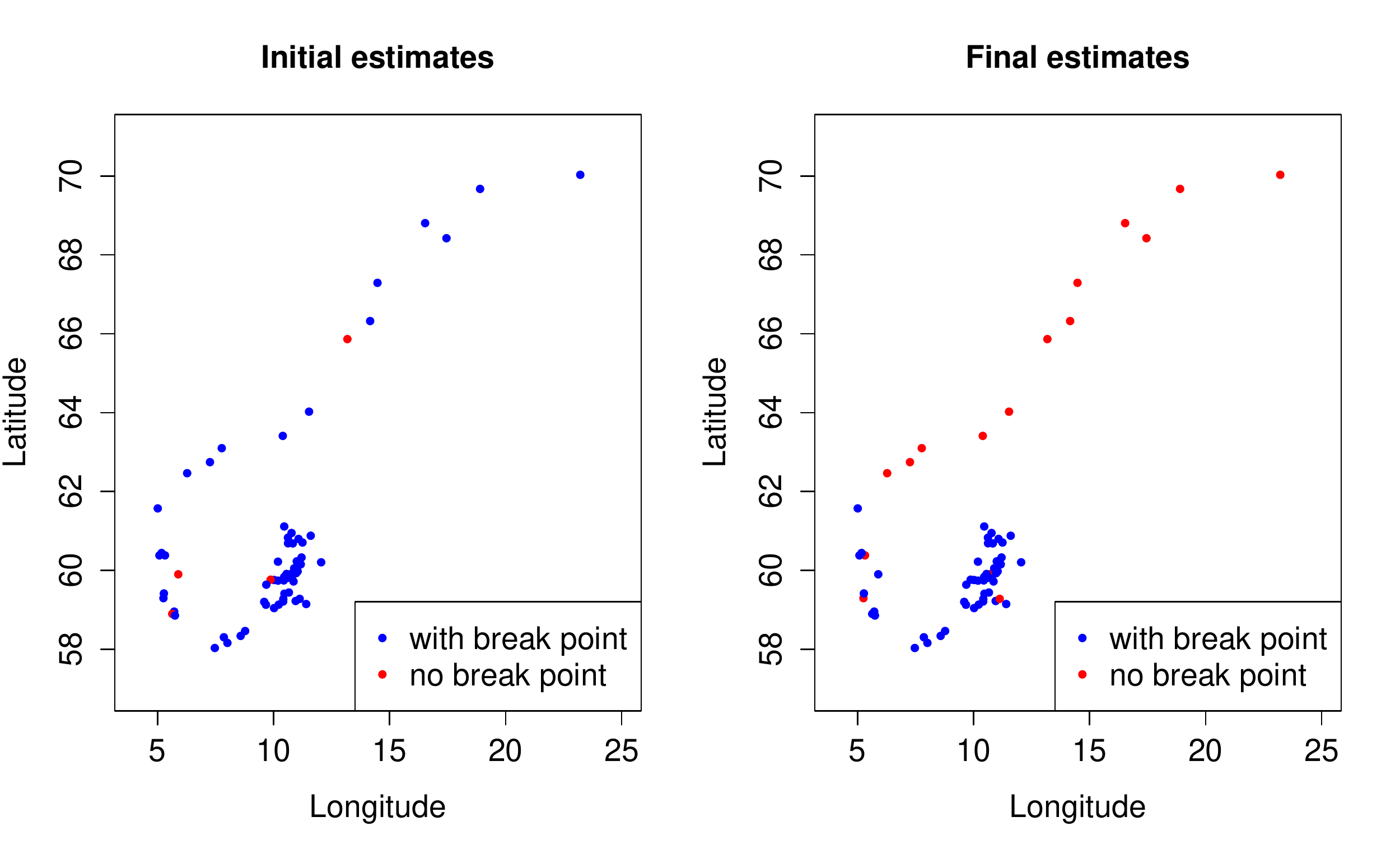}
		\caption{\label{fig:estimates}Initial and final estimates of the existence of the break points.}
	\end{figure}
	
	Second, We use the preliminary binary assignments of break points group based on BIC in a simple linear regression to assess whether there is variability in intercepts and population size effects. Both models with only main effects (indicator variable of group assignments, log population sizes, log distance-before/after break point) and those with main effects and interaction terms show evidence (p-value$<$0.05) of variability. Hence we apply the method shown in the simulation study for the analysis of the cell phone data. The difference in intercepts and population size effects is true both for the general population from all 427 counties, and for the user subpopulation we described above. 
	
\begin{figure}
		\centering
		\includegraphics[width=0.49\linewidth]{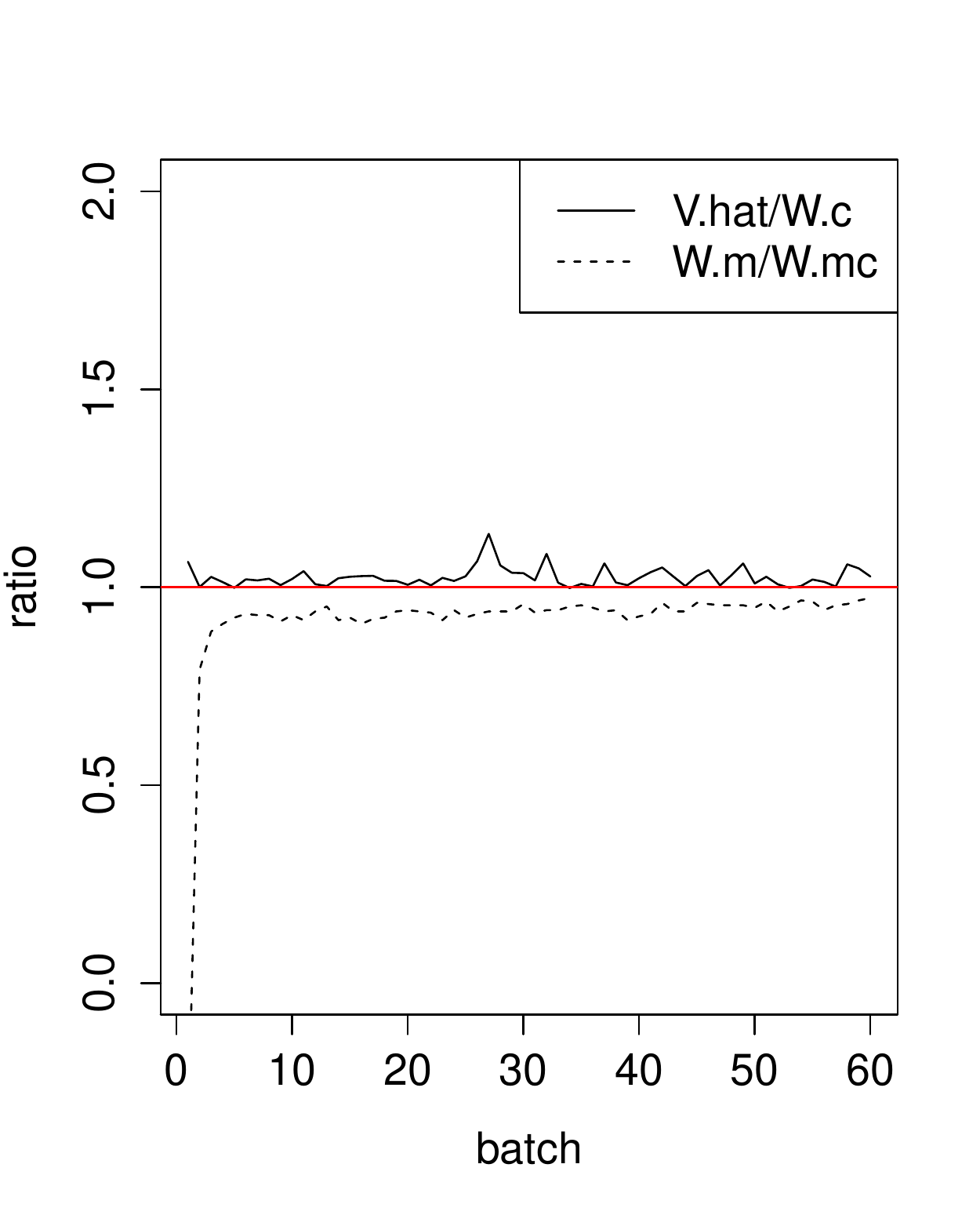}
		\includegraphics[width=0.49\linewidth]{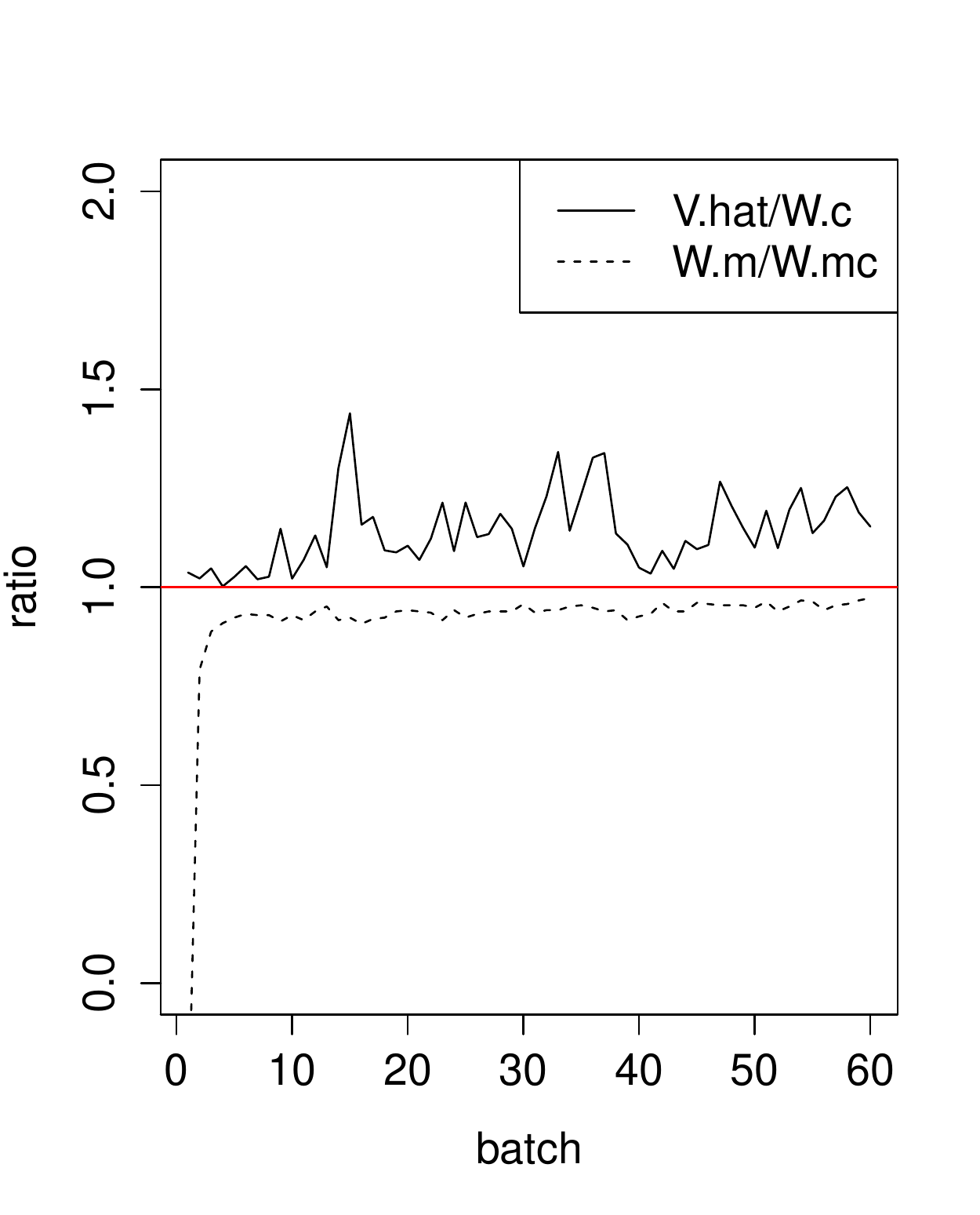}
		\caption{\label{fig:diag-1-}Left: diagnostic graph based on intercept estimates; Right: diagnostic graph based on $\sigma^2$; solid line: $\textrm{PSRF}_1$, dashed line: $\textrm{PSRF}_2$.}
	\end{figure}

	In the analysis of call records (CDRs) data (Figure \ref{slopes} and \ref{fig:estimates}), we observe that the slopes for source locations in the northeast appear to be less steep. Locations near the capital city, where the population is dense, are more likely to have breakpoints in the relationship of communication and distance . No such patterns were observed for slopes of other locations, both before and after the break points. Model estimates revealed that locations with no break point tend to be in the north while those with breakpoints are concentrated in the south. For diagnosis on convergence, Figure \ref{fig:diag-1-} shows a trend of $\textrm{PSRF}_2$ approaching 1 very quickly and a $\textrm{PSRF}_1$ fluctuating below 1.5, which is acceptable.
	
	\section{Discussion}
	To analyze the decline in communication intensity with geographical distance, we extended the gravity model by allowing for break-points in this relationship. We addressed the issue of the existence of break-points for each source location and quantify associated uncertainty using a Bayesian model. We also provided estimates of the slopes before and after each breakpoint. We investigated the geographical pattern of the existence of break-points and noted differences in these patterns between rural and urban areas.
	
	As an application of our method, we made use of an anonymized dataset of call detail records, using the number of mobile phone calls in our analysis as the measure of communication intensity between a pair of counties. The range of the outcomes is a count $\mathbb{N}$ before log transformation, and the regression model we specify treats the transformed outcomes as continuous, which is most appropriate when the number of calls between two locations is large (Figure \ref{fig:figure-2-}). In settings were there may be 0 or very low counts, one could consider alternative models (e.g. negative binomial) or the addition of an arbitrary small positive number to 0, although the latter approach can add bias \citep{flowerdew1982method, burger2009specification}. In this setting, a negative binomial model might be a better fit, though the interpretation of the parameters is less straightforward. Using Bayesian methods in a setting where the data are assumed to be negative binomial distributed requires non-standard approaches even without inclusion of break-points into models. \cite{zhou2012lognormal,pillow2012fully}, and \cite{polson2013bayesian} provide some useful tools for sequentially updating the parameters using Gibbs sampler by augmenting the posterior distribution with auxiliary parameters. When the number of counts is large, the negative binomial approach may not be computationally feasible; fitting negative binomial outcomes in Bayesian LASSO needs further investigation. One possible direction is to extend the methods based on the conditional normal distribution in \cite{polson2013bayesian} by transforming the variance matrix so that normal-distribution based LASSO method can be employed.
	
	Another extension of our methods would allow for aggregation of results across different subsamples; currently the number of locations we can analyze is limited by computational concerns. Developing a method to obtain consistent results from different overlapping sets of nodes-perhaps in a meta-analysis framework-- would alleviate the computational concerns, but is challenging. Some potentially useful approaches are provided by \cite{politis1994large,politis2001asymptotic,geyer20065601} and \cite{fitzenberger1998moving}. In particular, the stability selection in \cite{meinshausen2010stability} may be used to assess the properties of the meta-analytic results. An example of the use of LASSO in analyses that combine across subsamples arose from analyses intended to discover adverse drug reactions provided by \cite{ahmed2016class}. Another potentially useful approach is the use of path of partial posteriors in \cite{strathmann2015unbiased}. In this approach, the resampling procedure resembles the bootstrap, but with smaller resampling sizes. Because standard bootstrapping the LASSO estimator of the regression parameter for variance inference is known to yield inconsistent estimates \citep{knight2000asymptotics, chatterjee2010asymptotic}, modified bootstrapping must be used \citep{chatterjee2011bootstrapping}. Nonetheless, Bayesian LASSO procedures provide straightforward and valid estimates for standard errors.
	
	The findings from our analysis of mobile phone communication illustrate how such information might be used, should such communication networks prove to be accurate proxies for contact networks along which infectious diseases or other communicable processes spread. If so, such analyses might help guide designs of cluster randomized trials. Randomized trials ideally enroll participants in a way that minimizes the extent to which treatment assignment of one subject affects outcome of another. For interventions in which such interference occurs at the individual but not cluster level (e.g. through contacts among randomized subjects), cluster randomization can be useful \citep{campbell2007developments}. Clusters may be comprised of participants in the same geographical location, institution (e.g. school) or administrative unit (village). Cell phone data could potentially aid in the identification of appropriate clusters by providing information about the probability of interference.  When mixing across clusters cannot be eliminated, identification of treatment effects requires models of the mixing process \citep{carnegie2016estimation}. \cite{staples2015incorporating} and \cite{wang2014sample} investigated the impact of interference across randomized units on power of a clinical trial to detect effects of an intervention in preventing spread of infectious disease. As geographical distance is likely to affect contact networks, knowing the relationship between communication and distance may be useful not only for identification of clusters, but also to aid in development of appropriate mixing models.
	
	
	\section*{Conflict of interest}
	We declare no conflict of interest.

	
	\section*{Appendix}
	\subsection*{Computational complexity}
	
\begin{figure}
		\centering
		\includegraphics[width=0.8\linewidth]{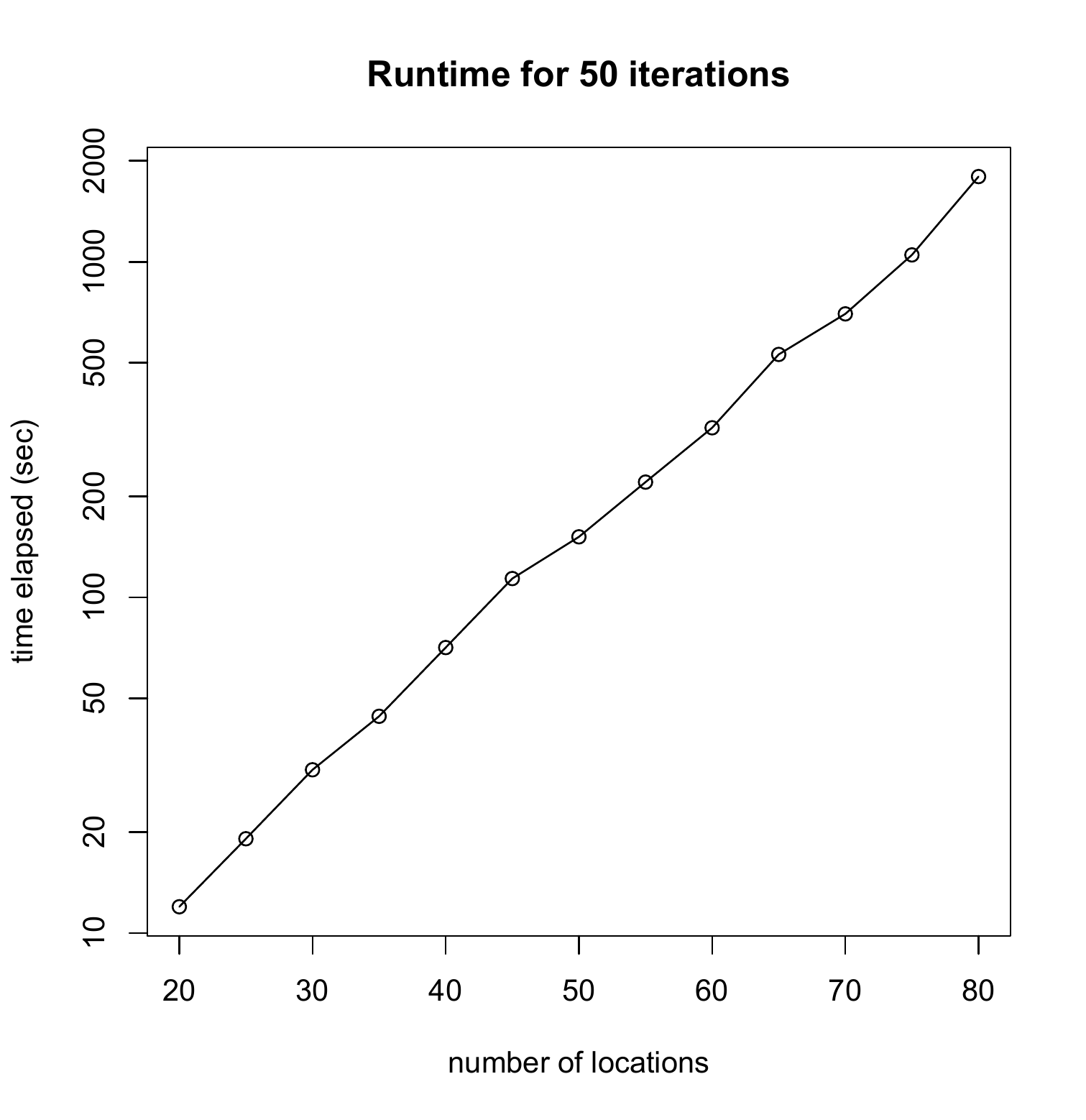}
		\caption{\label{fig:scaleup}Computation time (in seconds) versus number of locations in the simulation $s$. Note that the vertical axis is on logarithmic scale.}
	\end{figure}

	Please see Figure \ref{fig:scaleup}.
	
	\subsection*{Discussion on two models}
	Other models to study the impact of spatial distance on communication intensity have been proposed in the literature, such as the radiation model \citep{simini2012universal}, which predicts commuting fluxes between locations, and the rank-based friendship model proposed in \cite{liben2005geographic}, which ranks friendships based on the geographical distance between them. Both models reduce to Equation \ref{grav1} with certain constraints on their parameters or assumptions.
	The radiation model from \cite{simini2012universal} uses the following specification
	\begin{equation}
	\langle T_{ij} \rangle = T_i \dfrac{m_i n_j}{(m_i + s_{ij})(m_i + n_j + s_{ij})},
	\end{equation}
	where $\langle T_{ij} \rangle$ is the average commuting or mobility flux from location $i$ to $j$ (for simplicity, we denote average flux as $T_{ij}$ to keep the notation consistent), $T_i=\sum\limits_{j \neq i} T_{ij}$ is the total number of commuters from $i$. $s_{ij}$ is the population living in the circle centered at the source with a radius of $r_{ij}$ (not including $m_i$). Adopting this notation,
	\begin{equation}
	T_{ij} = T_i \dfrac{m_i n_j}{(m_i + s_{ij})(m_i + n_j + s_{ij})},
	\end{equation}
	
	Taking the logarithm yields,
	\begin{equation}
	\label{radi}
	\log(T_{ij}) = \log(T_i) + \log(m_i) + \log(n_j) - \log(m_i + s_{ij}) - \log(m_i + n_j + s_{ij}).
	\end{equation}
	
	As in \cite{simini2012universal}, we note that Equation \ref{radi} reduces to Equation \ref{grav1} with $\alpha+\beta=1$ and $\gamma=4$ when the population is uniformly distributed such that $m=n$ and $s_{ij} \approx m_i r_{ij}^2$. The model is mechanistic and has no parameter to fit.
	
	The rank-based friendship model in \cite{liben2005geographic} is formulated as follows. Let $u$ and $v$ be two individuals. They define $\text{rank}_u (v)=|\{w: d(u,w) <= d(u,v)\}|$, where $d(u,w)$ is the distance between individual $u$ and individual $w$. The probability of $u$ and $v$ being friends is modeled as
	\begin{equation}
	\text{Pr}[u \rightarrow u] \propto \dfrac{1}{\text{rank}_u(v)}.
	\label{rank}
	\end{equation}
	
	As $\text{rank}_u(v) \approx d(u,v)^2$ when the the population is uniformly distributed, Equation \ref{rank} reduces to Equation \ref{grav1} with $m=n=1$ and $\gamma=2$.
	
	Both the gravity and radiation models are based on strict assumptions of the underlying mechanism, which are hard to validate. The gravity model, which uses the same parameters for each pair of locations, implicitly assumes a homogeneous effect of distance for the intensity function. The radiation model addresses this issue by modeling the intrinsic heterogeneity of the geographical distribution of population as indicated by the incorporation of $s_{ij}$ in the model. However, subject to its strict assumption and `parameter-free' property, it does not allow for fluctuations of other forms or from other sources. The rank-based model deals with the heterogeneity by substituting distance with rank, which seems to have a similar role as the $s_{ij}$ in the radiation model. Thus the $\text{rank}$ function in Equation \ref{rank} can be regarded as an implicit function of distance and population distribution. We can make Equation \ref{rank} a parametric model by putting a parameter at the power of the rank, which when assuming population is uniformly distributed across the area, would be equivalent to the gravity model with parameter $\gamma$ for the distance $r_{ij}$.
	
	We note here that even though the rank-based approach shed some lights on the question in which we are interested, to move from resolution at individual level to zip code or county level requires a completely different set of assumptions. Therefore, a rank-based gravity model cannot be seen as a simple extension from the rank-based friendship model. 
	
	\bibliographystyle{apalike}
	\bibliography{distance}

\begin{thebibliography}{}

\bibitem[Ahmed et~al., 2016]{ahmed2016class}
Ahmed, I., Pariente, A., and Tubert-Bitter, P. (2016).
\newblock Class-imbalanced subsampling lasso algorithm for discovering adverse
  drug reactions.
\newblock {\em Statistical Methods in Medical Research}.

\bibitem[Balcan et~al., 2009]{balcan2009multiscale}
Balcan, D., Colizza, V., Gon{\c{c}}alves, B., Hu, H., Ramasco, J.~J., and
  Vespignani, A. (2009).
\newblock Multiscale mobility networks and the spatial spreading of infectious
  diseases.
\newblock {\em Proceedings of the National Academy of Sciences},
  106(51):21484--21489.

\bibitem[Brooks and Giudici, 1998]{brooks1998convergence}
Brooks, S. and Giudici, P. (1998).
\newblock Convergence assessment for reversible jump mcmc simulations.

\bibitem[Brooks and Gelman, 1998]{brooks1998general}
Brooks, S.~P. and Gelman, A. (1998).
\newblock General methods for monitoring convergence of iterative simulations.
\newblock {\em Journal of Computational and Graphical Statistics},
  7(4):434--455.

\bibitem[Buckee et~al., 2013]{buckee2013mobile}
Buckee, C.~O., Wesolowski, A., Eagle, N.~N., Hansen, E., and Snow, R.~W.
  (2013).
\newblock Mobile phones and malaria: modeling human and parasite travel.
\newblock {\em Travel Medicine and Infectious Disease}, 11(1):15--22.

\bibitem[Burger et~al., 2009]{burger2009specification}
Burger, M., Van~Oort, F., and Linders, G.-J. (2009).
\newblock On the specification of the gravity model of trade: zeros, excess
  zeros and zero-inflated estimation.
\newblock {\em Spatial Economic Analysis}, 4(2):167--190.

\bibitem[Campbell et~al., 2007]{campbell2007developments}
Campbell, M., Donner, A., and Klar, N. (2007).
\newblock Developments in cluster randomized trials and statistics in medicine.
\newblock {\em Statistics in Medicine}, 26(1):2--19.

\bibitem[Carnegie et~al., 2016]{carnegie2016estimation}
Carnegie, N.~B., Wang, R., and De~Gruttola, V. (2016).
\newblock Estimation of the overall treatment effect in the presence of
  interference in cluster-randomized trials of infectious disease prevention.
\newblock {\em Epidemiologic Methods}, 5(1):57--68.

\bibitem[Castelloe and Zimmerman, 2002]{castelloe2002convergence}
Castelloe, J.~M. and Zimmerman, D.~L. (2002).
\newblock Convergence assessment for reversible jump mcmc samplers.
\newblock {\em Department of Statistics and Actuarial Science, University of
  Iowa, Technical Report}, 313.

\bibitem[Chatterjee and Lahiri, 2010]{chatterjee2010asymptotic}
Chatterjee, A. and Lahiri, S. (2010).
\newblock Asymptotic properties of the residual bootstrap for lasso estimators.
\newblock {\em Proceedings of the American Mathematical Society},
  138(12):4497--4509.

\bibitem[Chatterjee and Lahiri, 2011]{chatterjee2011bootstrapping}
Chatterjee, A. and Lahiri, S.~N. (2011).
\newblock Bootstrapping lasso estimators.
\newblock {\em Journal of the American Statistical Association},
  106(494):608--625.

\bibitem[Cs{\'a}ji et~al., 2013]{csaji2013exploring}
Cs{\'a}ji, B.~C., Browet, A., Traag, V.~A., Delvenne, J.-C., Huens, E.,
  Van~Dooren, P., Smoreda, Z., and Blondel, V.~D. (2013).
\newblock Exploring the mobility of mobile phone users.
\newblock {\em Physica A: Statistical Mechanics and its Applications},
  392(6):1459--1473.

\bibitem[Eagle et~al., 2008]{eagle2008mobile}
Eagle, N., Pentland, A.~S., and Lazer, D. (2008).
\newblock Mobile phone data for inferring social network structure.
\newblock In {\em Social Computing, Behavioral Modeling, and Prediction}, pages
  79--88. Springer.

\bibitem[Eagle et~al., 2009]{eagle2009inferring}
Eagle, N., Pentland, A.~S., and Lazer, D. (2009).
\newblock Inferring friendship network structure by using mobile phone data.
\newblock {\em Proceedings of the National Academy of Sciences},
  106(36):15274--15278.

\bibitem[Efron et~al., 2004]{efron2004least}
Efron, B., Hastie, T., Johnstone, I., Tibshirani, R., et~al. (2004).
\newblock Least angle regression.
\newblock {\em The Annals of statistics}, 32(2):407--499.

\bibitem[Expert et~al., 2011]{expert2011uncovering}
Expert, P., Evans, T.~S., Blondel, V.~D., and Lambiotte, R. (2011).
\newblock Uncovering space-independent communities in spatial networks.
\newblock {\em Proceedings of the National Academy of Sciences},
  108(19):7663--7668.

\bibitem[Fitzenberger, 1998]{fitzenberger1998moving}
Fitzenberger, B. (1998).
\newblock The moving blocks bootstrap and robust inference for linear least
  squares and quantile regressions.
\newblock {\em Journal of Econometrics}, 82(2):235--287.

\bibitem[Flowerdew and Aitkin, 1982]{flowerdew1982method}
Flowerdew, R. and Aitkin, M. (1982).
\newblock A method of fitting the gravity model based on the poisson
  distribution.
\newblock {\em Journal of Regional Science}, 22(2):191--202.

\bibitem[Gelman et~al., 2014]{gelman2014bayesian}
Gelman, A., Carlin, J.~B., Stern, H.~S., and Rubin, D.~B. (2014).
\newblock {\em Bayesian data analysis}, volume~2.
\newblock Taylor \& Francis.

\bibitem[Gelman and Rubin, 1992]{gelman1992inference}
Gelman, A. and Rubin, D.~B. (1992).
\newblock Inference from iterative simulation using multiple sequences.
\newblock {\em Statistical Science}, pages 457--472.

\bibitem[Geyer, 2006]{geyer20065601}
Geyer, C.~J. (2006).
\newblock 5601 notes: The subsampling bootstrap.
\newblock {\em Unpublished manuscript}.

\bibitem[Granovetter, 1973]{granovetter1973strength}
Granovetter, M.~S. (1973).
\newblock The strength of weak ties.
\newblock {\em American Journal of Sociology}, 78(6):1360--1380.

\bibitem[Green, 1995]{green1995reversible}
Green, P.~J. (1995).
\newblock Reversible jump markov chain monte carlo computation and bayesian
  model determination.
\newblock {\em Biometrika}, 82(4):711--732.

\bibitem[Green and Hastie, 2009]{green2009reversible}
Green, P.~J. and Hastie, D.~I. (2009).
\newblock Reversible jump mcmc.
\newblock {\em Genetics}, 155(3):1391--1403.

\bibitem[Gregson et~al., 2002]{gregson2002sexual}
Gregson, S., Nyamukapa, C.~A., Garnett, G.~P., Mason, P.~R., Zhuwau, T.,
  Cara{\"e}l, M., Chandiwana, S.~K., and Anderson, R.~M. (2002).
\newblock Sexual mixing patterns and sex-differentials in teenage exposure to
  hiv infection in rural zimbabwe.
\newblock {\em The Lancet}, 359(9321):1896--1903.

\bibitem[Hawelka et~al., 2014]{hawelka2014geo}
Hawelka, B., Sitko, I., Beinat, E., Sobolevsky, S., Kazakopoulos, P., and
  Ratti, C. (2014).
\newblock Geo-located twitter as proxy for global mobility patterns.
\newblock {\em Cartography and Geographic Information Science}, 41(3):260--271.

\bibitem[Helleringer and Kohler, 2007]{helleringer2007sexual}
Helleringer, S. and Kohler, H.-P. (2007).
\newblock Sexual network structure and the spread of hiv in africa: evidence
  from likoma island, malawi.
\newblock {\em AIDS}, 21(17):2323--2332.

\bibitem[Jones and Handcock, 2003]{jones2003assessment}
Jones, J.~H. and Handcock, M.~S. (2003).
\newblock An assessment of preferential attachment as a mechanism for human
  sexual network formation.
\newblock {\em Proceedings of the Royal Society of London B: Biological
  Sciences}, 270(1520):1123--1128.

\bibitem[Knight and Fu, 2000]{knight2000asymptotics}
Knight, K. and Fu, W. (2000).
\newblock Asymptotics for lasso-type estimators.
\newblock {\em Annals of Statistics}, pages 1356--1378.

\bibitem[Krings et~al., 2009]{krings2009urban}
Krings, G., Calabrese, F., Ratti, C., and Blondel, V.~D. (2009).
\newblock Urban gravity: a model for inter-city telecommunication flows.
\newblock {\em Journal of Statistical Mechanics: Theory and Experiment},
  2009(07):L07003.

\bibitem[Lambiotte et~al., 2008]{lambiotte2008geographical}
Lambiotte, R., Blondel, V.~D., de~Kerchove, C., Huens, E., Prieur, C., Smoreda,
  Z., and Van~Dooren, P. (2008).
\newblock Geographical dispersal of mobile communication networks.
\newblock {\em Physica A: Statistical Mechanics and its Applications},
  387(21):5317--5325.

\bibitem[Liben-Nowell et~al., 2005]{liben2005geographic}
Liben-Nowell, D., Novak, J., Kumar, R., Raghavan, P., and Tomkins, A. (2005).
\newblock Geographic routing in social networks.
\newblock {\em Proceedings of the National Academy of Sciences},
  102(33):11623--11628.

\bibitem[Meinshausen and B{\"u}hlmann, 2010]{meinshausen2010stability}
Meinshausen, N. and B{\"u}hlmann, P. (2010).
\newblock Stability selection.
\newblock {\em Journal of the Royal Statistical Society: Series B (Statistical
  Methodology)}, 72(4):417--473.

\bibitem[Noulas et~al., 2012]{noulas2012tale}
Noulas, A., Scellato, S., Lambiotte, R., Pontil, M., and Mascolo, C. (2012).
\newblock A tale of many cities: universal patterns in human urban mobility.
\newblock {\em PloS one}, 7(5):e37027.

\bibitem[Onnela et~al., 2011]{onnela2011geographic}
Onnela, J.-P., Arbesman, S., Gonz{\'a}lez, M.~C., Barab{\'a}si, A.-L., and
  Christakis, N.~A. (2011).
\newblock Geographic constraints on social network groups.
\newblock {\em PLoS one}, 6(4):e16939.

\bibitem[Onnela et~al., 2007]{onnela2007structure}
Onnela, J.-P., Saram{\"a}ki, J., Hyv{\"o}nen, J., Szab{\'o}, G., Lazer, D.,
  Kaski, K., Kert{\'e}sz, J., and Barab{\'a}si, A.-L. (2007).
\newblock Structure and tie strengths in mobile communication networks.
\newblock {\em Proceedings of the National Academy of Sciences},
  104(18):7332--7336.

\bibitem[Park and Casella, 2008]{park2008bayesian}
Park, T. and Casella, G. (2008).
\newblock The bayesian lasso.
\newblock {\em Journal of the American Statistical Association},
  103(482):681--686.

\bibitem[Pillow and Scott, 2012]{pillow2012fully}
Pillow, J.~W. and Scott, J.~G. (2012).
\newblock Fully bayesian inference for neural models with negative-binomial
  spiking.
\newblock In {\em NIPS}, pages 1907--1915.

\bibitem[Politis and Romano, 1994]{politis1994large}
Politis, D.~N. and Romano, J.~P. (1994).
\newblock Large sample confidence regions based on subsamples under minimal
  assumptions.
\newblock {\em The Annals of Statistics}, pages 2031--2050.

\bibitem[Politis et~al., 2001]{politis2001asymptotic}
Politis, D.~N., Romano, J.~P., and Wolf, M. (2001).
\newblock On the asymptotic theory of subsampling.
\newblock {\em Statistica Sinica}, pages 1105--1124.

\bibitem[Polson et~al., 2013]{polson2013bayesian}
Polson, N.~G., Scott, J.~G., and Windle, J. (2013).
\newblock Bayesian inference for logistic models using p{\'o}lya--gamma latent
  variables.
\newblock {\em Journal of the American statistical Association},
  108(504):1339--1349.

\bibitem[Porter et~al., 2009]{porter2009communities}
Porter, M.~A., Onnela, J.-P., and Mucha, P.~J. (2009).
\newblock Communities in networks.
\newblock {\em Notices of the AMS}, 56(9):1082--1097.

\bibitem[Sailer and McCulloh, 2012]{sailer2012social}
Sailer, K. and McCulloh, I. (2012).
\newblock Social networks and spatial configuration—how office layouts drive
  social interaction.
\newblock {\em Social Networks}, 34(1):47--58.

\bibitem[Simini et~al., 2012]{simini2012universal}
Simini, F., Gonz{\'a}lez, M.~C., Maritan, A., and Barab{\'a}si, A.-L. (2012).
\newblock A universal model for mobility and migration patterns.
\newblock {\em Nature}, 484(7392):96--100.

\bibitem[Staples et~al., 2015]{staples2015incorporating}
Staples, P.~C., Ogburn, E.~L., and Onnela, J.-P. (2015).
\newblock Incorporating contact network structure in cluster randomized trials.
\newblock {\em Scientific Reports}, 5.

\bibitem[Strathmann et~al., 2015]{strathmann2015unbiased}
Strathmann, H., Sejdinovic, D., and Girolami, M. (2015).
\newblock Unbiased bayes for big data: Paths of partial posteriors.
\newblock {\em arXiv preprint arXiv:1501.03326}.

\bibitem[Tatem et~al., 2014]{tatem2014integrating}
Tatem, A.~J., Huang, Z., Narib, C., Kumar, U., Kandula, D., Pindolia, D.~K.,
  Smith, D.~L., Cohen, J.~M., Graupe, B., Uusiku, P., et~al. (2014).
\newblock Integrating rapid risk mapping and mobile phone call record data for
  strategic malaria elimination planning.
\newblock {\em Malaria journal}, 13(1):52.

\bibitem[Ter~Wal and Boschma, 2009]{ter2009applying}
Ter~Wal, A.~L. and Boschma, R.~A. (2009).
\newblock Applying social network analysis in economic geography: framing some
  key analytic issues.
\newblock {\em The Annals of Regional Science}, 43(3):739--756.

\bibitem[Tibshirani, 1996]{tibshirani1996regression}
Tibshirani, R. (1996).
\newblock Regression shrinkage and selection via the lasso.
\newblock {\em Journal of the Royal Statistical Society. Series B
  (Methodological)}, pages 267--288.

\bibitem[Wang et~al., 2011]{wang2011human}
Wang, D., Pedreschi, D., Song, C., Giannotti, F., and Barabasi, A.-L. (2011).
\newblock Human mobility, social ties, and link prediction.
\newblock In {\em Proceedings of the 17th ACM SIGKDD international conference
  on Knowledge discovery and data mining}, pages 1100--1108. ACM.

\bibitem[Wang et~al., 2013]{wang2013human}
Wang, L., Wang, Z., Zhang, Y., and Li, X. (2013).
\newblock How human location-specific contact patterns impact spatial
  transmission between populations?
\newblock {\em Scientific Reports}, 3.

\bibitem[Wang et~al., 2014]{wang2014sample}
Wang, R., Goyal, R., Lei, Q., Essex, M., and De~Gruttola, V. (2014).
\newblock Sample size considerations in the design of cluster randomized trials
  of combination hiv prevention.
\newblock {\em Clinical Trials}, 11(3):309--318.

\bibitem[Zhou et~al., 2012]{zhou2012lognormal}
Zhou, M., Li, L., Dunson, D., and Carin, L. (2012).
\newblock Lognormal and gamma mixed negative binomial regression.
\newblock In {\em Machine learning: proceedings of the International
  Conference. International Conference on Machine Learning}, volume 2012, page
  1343. NIH Public Access.

\end{thebibliography}

	%
	%
	%
	%
	
\end{document}